\documentclass[apjl]{emulateapj}
\def\apjl{ApJL }
\def\aj{AJ }
\def\apj{ApJ }
\def\pasp{PASP }

\def\apjs{ApJS }

\def\aap{A\&A }

\usepackage{hyperref}
\hypersetup{colorlinks=true,citecolor=blue,linkcolor=black,filecolor=black,runcolor=black}
 \usepackage{natbib}
\shorttitle{The GALEX/S$^{4}$G UV-IR color-color diagram}
\shortauthors{Bouquin et al.}
\begin{document}

\title{The GALEX/S$^{4}$G UV-IR color-color diagram: Catching spiral galaxies away from the Blue Sequence}

\author{Alexandre Y. K. Bouquin\altaffilmark{1}, Armando Gil de Paz\altaffilmark{1}, Samuel Boissier\altaffilmark{2,3}, Juan-Carlos Mu\~noz-Mateos\altaffilmark{4}, Kartik Sheth\altaffilmark{5}, Dennis Zaritsky\altaffilmark{6}, Jarkko Laine\altaffilmark{7}, Jes\'{u}s Gallego\altaffilmark{1}, Reynier F. Peletier\altaffilmark{8}, Benjamin R. R\"{o}ck\altaffilmark{9,10}, Johan H. Knapen\altaffilmark{9,10}} 
%S\'{e}bastien Comer\'{o}n\altaffilmark{5}, Ron Buta\altaffilmark{6}, Jaime Zamorano\altaffilmark{1}, }

\affil{\altaffilmark{1}Departamento de Astrof\'isica y CC. de la Atm\'osfera, Universidad Complutense de Madrid, E-28040 Madrid, Spain}
%\email{abouquin@fis.ucm.es}
%\email{agil@fis.ucm.es}
%\email{j.gallego@fis.ucm.es}
%\email{jzamorano@fis.ucm.es}
\affil{\altaffilmark{2}Aix Marseille Universit\'e, CNGRS, LAM (Laboratoire d'Astrophysique de Marseille) UMR 7326, 13388, Marseille, France}
\affil{\altaffilmark{3}I.N.A.F., Osservatorio Astronomico di Bologna, via Ranzani 1 I-40127 Bologna, Italy}
%\email{samuel.boissier@oamp.fr}
\affil{\altaffilmark{4}European Southern Observatory, Casilla 19001, Santiago 19, Chile}
%\email{jmunoz@nrao.edu}
\affil{\altaffilmark{5}National Radio Astronomy Observatory/NAASC, 520 Edgemont Road, Charlottesville, VA 22903, USA}
%\email{astrokartik@gmail.com}
\affil{\altaffilmark{6}Steward Observatory, University of Arizona, 933 North Cherry Avenue, Tucson, AZ 85721, USA}
%\email{dennis.zaritsky@gmail.com}
\affil{\altaffilmark{7}Astronomy Division, Department of Physics, FIN-90014 University of Oulu, P.O. Box 3000, Oulu, Finland}
%\email{jarkko.laine@oulu.fi}
\affil{\altaffilmark{8}Kapteyn Astronomical Institute, Postbus 800, 9700 AV Groningen, the Netherlands}
%\email{peletier@astro.rug.nl}
\affil{\altaffilmark{9}Instituto de Astrof\'isica de Canarias, V\'ia L\'actea, S/N, 38205 La Laguna, Spain}
%\email{jhk@iac.es}
\affil{\altaffilmark{10}Departamento de Astrof\'isica, Universidad de La Laguna, E-38205 La Laguna, Tenerife, Spain}

%\email{seb.comeron@gmail.com}
%\affil{\altaffilmark{6}Department of Physics and Astronomy, University of Alabama, Box 870324, Tuscaloosa, AL 35487}
%\email{buta@sarah.astr.ua.edu}

\begin{abstract}
We obtained GALEX FUV, NUV, and Spitzer/IRAC 3.6\,$\mu$m photometry for $>$2000 galaxies, available for 90\% of the S$^{4}$G sample. We find a very tight ``GALEX Blue Sequence (GBS)" in the ($\textit{FUV}$\,$-$\,$\textit{NUV}$) versus ($\textit{NUV}$\,$-$\,[3.6]) color-color diagram which is populated by irregular and spiral galaxies, and is mainly driven by changes in the formation timescale ($\tau$) and a degeneracy between $\tau$ and dust reddening. The tightness of the GBS provides an unprecedented way of identifying star-forming galaxies  and objects that are just evolving to (or from) what we call the ``GALEX Green Valley (GGV)". At the red end of the GBS, at ($\textit{NUV}$\,$-$\,[3.6])\,$>$\,5, we find a wider ``GALEX Red Sequence (GRS)" mostly populated by E/S0 galaxies that has a perpendicular slope to that of the GBS and of the optical red sequence. We find no such dichotomy in terms of stellar mass (measured by $\textit{M}$$_{[3.6]}$), since both massive ($M_{\star}$\,$>$\,$10^{11}$\,$M_{\odot}$) blue and red sequence galaxies are identified. The type that is proportionally more often found in the GGV are the S0-Sa's and most of these are ​located​ in high-density environments. ​​​We discuss evolutionary models of galaxies that show a rapid transition from the blue to the red sequence on timescale of $10^{8}$\,years.
\end{abstract}

\section{INTRODUCTION}
Color-magnitude and color-color diagrams are proven galaxy evolution diagnostic tools. Recent studies making use of $\textit{GALEX}$ and SDSS data, such as \citet{Wyder2007}, have clearly shown the existence of a bimodal distribution of galaxies in ($\textit{NUV}$\,$-$\,$\textit{r}$) vs $\textit{r}$ color-magnitude diagrams (CMDs), where the redder and brighter region is populated mainly by early-type galaxies in a ``red sequence", and where the bluer and fainter part of the diagram is populated mainly by late-type galaxies in a ``blue sequence".

We construct a color-magnitude and a color-color diagram using far-ultraviolet ($\textit{FUV}$; effective wavelength 151.6\,nm), near-ultraviolet ($\textit{NUV}$; 226.7\,nm) images from the $\textit{Galaxy Evolution Explorer}$ \citep[$\textit{GALEX}$;][]{Martin2005}, and 3.6\,micron images from the images released by the \textit{Spitzer} Survey of Stellar Structure in Galaxies \citep[S$^{4}$G,][]{Sheth2010}. 
The near-infrared (NIR) band is sensitive mainly to old stars, and the UV bands to young stars ($<$1\,Gyr). As we show below, a combination of UV and NIR bands allows us to separate star-forming from passively-evolving galaxies, while the use of the two $\textit{GALEX}$ UV bands makes this analysis very sensitive to transitional galaxies and to the transition timescale, as the $\textit{FUV}$ band is most sensitive to the presence of young (OB-type) stars. Our sample is based on the $\textit{Spitzer}$ Survey of Stellar Structure in Galaxies \citep[S$^{4}$G;][]{Sheth2010}. In this study, we gather the publicly available $\textit{GALEX}$ data (from data release GR6/7) for these S$^{4}$G galaxies, and measure $\textit{NUV}$ and $\textit{FUV}$ surface and asymptotic photometry for over 2000 galaxies ($\sim$90\% of the S$^{4}$G galaxies). We then combine this with consistent surface and asymptotic 3.6$\mu$m photometry from S$^{4}$G.
We adopt a value of $\textit{H}$$_{0}$\,=\,75\,km\,s$^{-1}$\,Mpc$^{-1}$ for galaxies lacking a redshift-independent distance measurement \citep[as in][]{Sheth2010}.

\section{SAMPLE}
\begin{figure*}
\begin{center}
\includegraphics[width=1\textwidth]{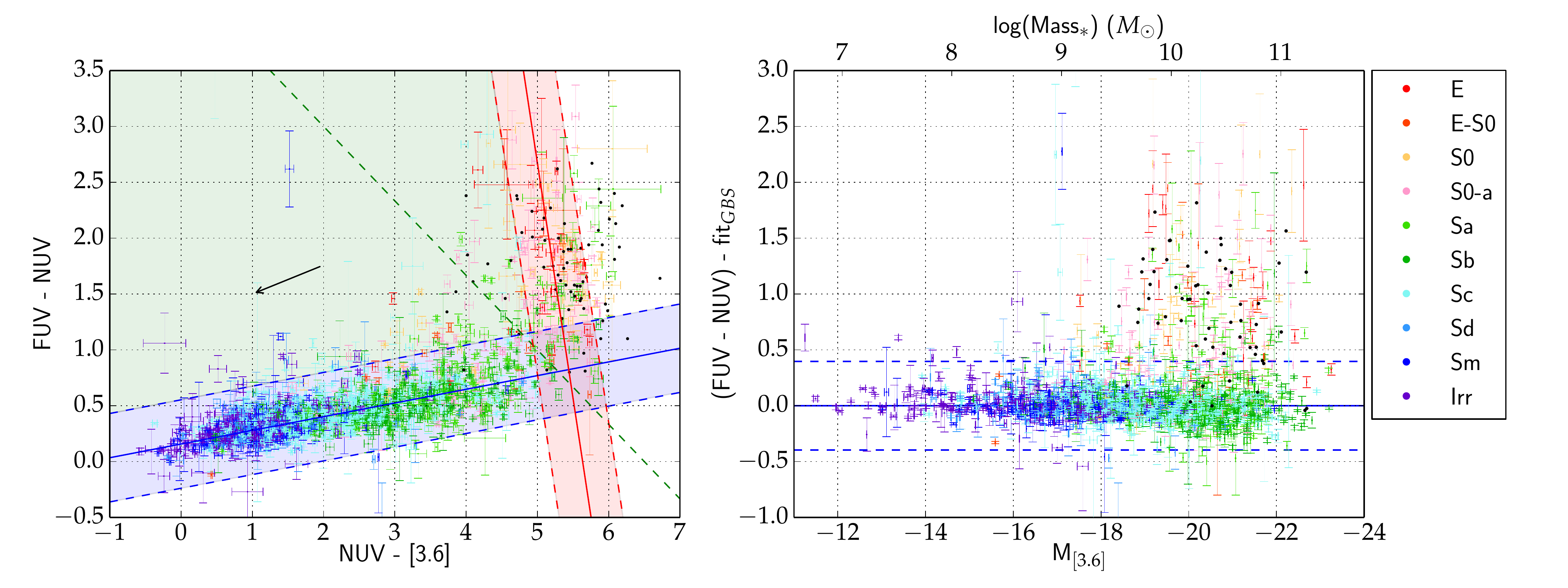}
\caption{\textit{Left}\,: ($\textit{FUV}$\,$-$\,$\textit{NUV}$) versus ($\textit{NUV}$\,$-$\,[3.6]) color-color diagram. Morphological types are represented by color (see legend). We define the blue and red sequence by dividing the plot into two regions and fitting a least-squares 1-D polynomial (a line) to each region. The solid blue line is the weighted-by-error linear fit for the lower-left region, and the solid red line is the weighted-by-error linear fit for the upper-right region. The dashed lines are parallel to their solid counterpart, but translated in the ($\textit{FUV}$\,$-$\,$\textit{NUV}$) direction by $\pm$2 sigma for the GBS (blue shaded region), and translated in the ($\textit{NUV}$\,$-$\,[3.6]) direction by $\pm$1 sigma for the GRS (red shaded region). The GGV is also shown (green shaded region). The black arrow indicates the reddening vector for an attenuation of A$_{V}$\,=\,0.5\,mag. The black dots are measurements for the \citet{Cappellari2013} sample (section \ref{UVIRCMD}; not included in the statistics) obtained by \citet{Zaritsky2015a} (we adopted here ($\textit{I}$\,$-$\,[3.6])=0.5\,mag). We use a 3.6\,$\mu$m stellar mass-to-light ratio $\Upsilon_{3.6}$\,=\,0.6 or $\rm{log}(\Upsilon_{3.6}$)\,=\,-0.22 \citep[IMF: Chabrier;][]{Meidt2014}. \textit{Right}\,: Color-magnitude diagram using the difference between ($\textit{FUV}$\,$-$\,$\textit{NUV}$) color and the fit to the GBS in the color-color diagram versus M$_{[3.6]}$. Same color-coding as in the left panel. \label{fig:ColorColorDiagram}}
\end{center}
\end{figure*}
The S$^{4}$G is a volume-limited (\textit{d}\,$<$\,40\,Mpc), magnitude-limited (m$_{\rm{Bcorr}}$\,$<$\,15.5\,mag, corrected for inclination, galactic extinction, and $\textit{K}$-correction), and size-limited ($D_{25}$\,$>$\,1\,$\arcmin$) survey avoiding the Galactic equatorial plane($\vert b \vert$\,$>$\,30\,$\arcdeg$) of 2,352 galaxies, consisting of 3.6\,$\mu$m and 4.5\,$\mu$m IRAC-band images.
Galaxies covered by this survey account for the majority of the significantly-sized galaxies in our extragalactic neighborhood, including galaxies of all Hubble types and masses, in various environments from field to clusters of galaxies (including 187 Virgo galaxies). The HI 21\,cm-redshift selection, however, biases the sample against very massive ETGs (Early-Type Galaxies)\footnote{An extension of the S$^{4}$G sample, aimed to correct for the HI selection effect, is currently in progress.} \citep{Lagos2014b}.
We compiled a subsample of corresponding $\textit{GALEX}$ (data release GR6/7) tiles using GALEXView\footnote{http://galex.stsci.edu/GalexView/}, in both $\textit{FUV}$ and $\textit{NUV}$ for over 2,100 galaxies.

$\textit{FUV}$ and $\textit{NUV}$ asymptotic magnitudes were obtained using the same method as in \citet{gdp2007b}, corrected for the Milky Way foreground attenuation, measured for 2053 galaxies. After excluding shallow observations with magnitude errors larger than unity, and bad photometry due to contamination from bright stars, we are left with 1931 galaxies. The S$^{4}$G 3.6$\mu$m photometry has been obtained using an identical procedure but adapted to the specifics of IRAC \citep{Munoz2015}. All magnitudes throughout this paper are given in the AB system.
Numerical morphological types, absolute $\textit{B}$-band magnitudes and optical colors were obtained from HyperLeda \citep{Paturel2003}. The total UV coverage of our parent S$^{4}$G sample is \,$\sim$\,90\%, with a rather uniform fraction across all properties including absolute $\textit{B}$-band magnitude, distance, and morphological types, and therefore represents and inherits the selection criteria of S$^{4}$G. 

\section{RESULTS AND ANALYSIS}
\begin{figure*}
\begin{center}
\includegraphics[width=1\textwidth]{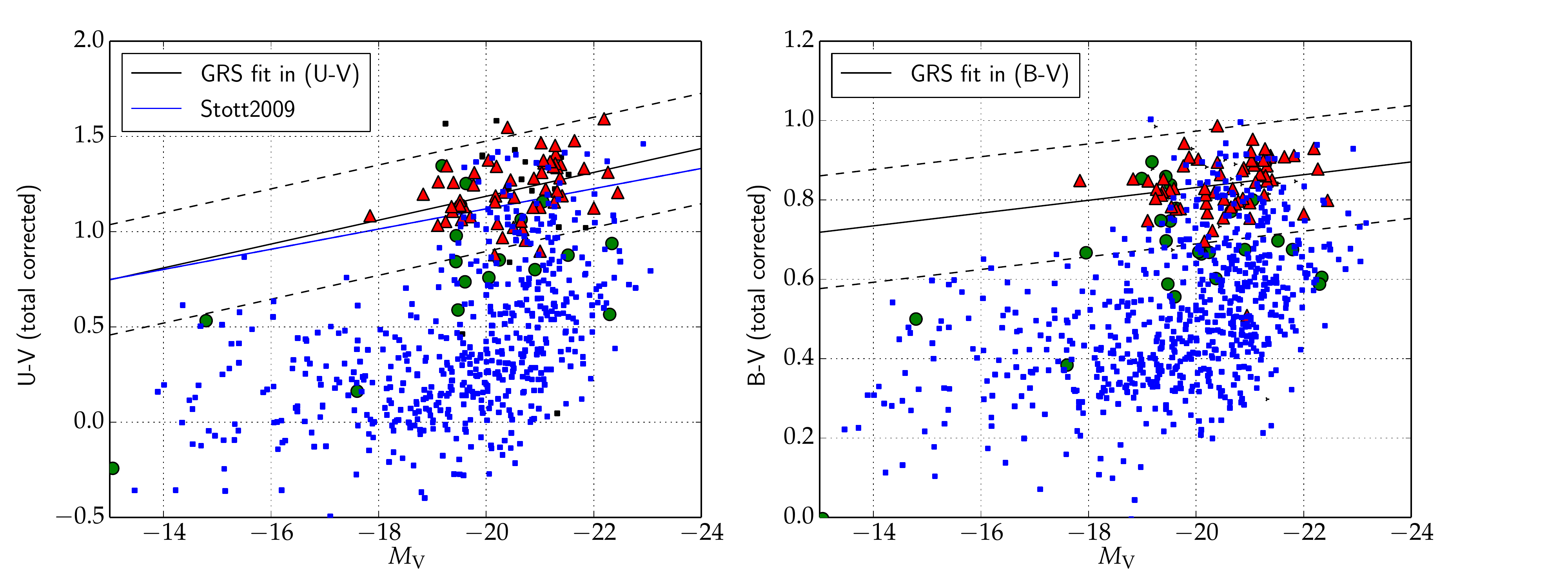}
\caption{GBS ($\textit{blue squares}$), GGV ($\textit{green points}$), and GRS ($\textit{red triangles}$) galaxies are shown, with a linear fit to the GRS galaxies (solid lines) and the $rms_{\rm{GRS}}$\,=\,2 (dashed lines). $\textit{Left}$\,: Optical CMD ($U-V$) versus $V$-band absolute magnitude of our sample. The solid blue line is the red sequence fit from \citet{Stott2009} with the intercept value adjusted for comparison. $\textit{Right}$\,: ($B-V$) versus $V$-band absolute magnitude. \label{fig:Optical-IR-CMD}}
\end{center}
\end{figure*}
\subsection{UV-IR Color-Color Diagram}
Using the asymptotic photometry described above, we construct the ($\textit{FUV}$\,$-$\,$\textit{NUV}$) versus ($\textit{NUV}$\,$-$\,[3.6]) color-color diagram (Fig.\ref{fig:ColorColorDiagram}). We find a clear bimodal pattern: a very tight blue sequence, which we call the ``$\textit{GALEX}$ Blue Sequence'' (GBS hereafter), and a less tight red sequence we call the ``$\textit{GALEX}$ Red sequence'' (GRS hereafter). The distribution of the GRS is perpendicular to that seen in optical color-color diagrams \citep[e.g.,][]{Strateva2001} likely due to the mass and metallicity dependence of the UV upturn \citep{Boselli2005}. The more massive and metal-rich a galaxy is, the more important the contribution from the UV upturn to the UV emission and bluer the ($\textit{FUV}$\,$-$\,$\textit{NUV}$) color. Alternative use of this color are still being explored \citep*[e.g.,][]{HernandezPerez2014,Zaritsky2014,Zaritsky2015a}.
\\

To quantify the GBS and the GRS, we split the color-color plane by choosing a line that passes through coordinates (2,3) and (5,1) (left panel of  Fig.\ref{fig:ColorColorDiagram}).
We fit the bluer-half with a simple error-weighted linear regression (the subsample size here is 1,787 galaxies), giving us a slope of 0.12$\pm$0.01 and a $\textit{y}$-intercept at 0.16$\pm$0.01. The standard deviation of the sample over this range is $\textit{rms}$$_{\rm{GBS}}$\,=\,0.20 while a Gaussian fit to the residuals yields an even smaller value of $\sigma_{\rm{GBS}}$\,=\,0.10.

We also fit the redder-half (144 galaxies) with a simple error-weighted linear regression (in the ($\textit{NUV}$\,$-$\,[3.6]) direction), giving us a slope of $-$4.21$\pm$1.62, and a $\textit{rms}$$_{\rm{GRS}}$\,=\,0.45.
The location of galaxies in this diagram is strongly correlated with the Hubble type (see also Fig.\ref{fig:nine}).
There are only two true ellipticals in the GBS, ESO548-023 and NGC 855. Some of the ellipticals found in the GBS are actually blue compact dwarfs (BCDs) morphologically misclassified as ellipticals. The latter plus the 31 later-typed ETGs (E-S0,S0) located in the GBS could be either in the tail of star formation \citep{Bresolin2013} or could be rejuvenated systems \citep{Kannappan2009,Thilker2010}. Finally, early-type spiral galaxies (S0-a, Sa, Sab) are populating the GBS, the GRS and the region in between (which we name $\textit{GALEX}$ Green Valley; GGV hereafter) which is consistent with a scenario where these galaxies are transitional objects. 

\subsection{UV-IR Color-Magnitude Diagram} \label{UVIRCMD}
Fig.\ref{fig:ColorColorDiagram} (Right) shows the residuals of the UV color to the GBS linear fit obtained in our UV-IR color-color diagram (Left), ($\textit{FUV}$\,$-$\,$\textit{NUV}$)\,$-$\,($\textit{FUV}$\,$-$\,$\textit{NUV}$)$_{\rm{GBS}}$, as a fraction of the absolute magnitude at 3.6\,$\mu$m, $\textit{M}$$_{[3.6]}$ and stellar mass measured assuming a $\Upsilon_{3.6}$ of 0.6\,$M_{\odot}$/$L_{\odot}$ \citep{Meidt2014,Rock2015}. Stellar mass is not the driver for the evolution of these systems in or out the GBS. The distribution of our samples of ETGs \citep[our data and those from][]{Zaritsky2015a} is similar, meaning that the bias against HI-poor ETGs in S$^{4}$G is not driving our results.  

\subsection{Optical Color-Magnitude Diagram}
\begin{figure*}
\begin{center}
\includegraphics[width=1.0\textwidth]{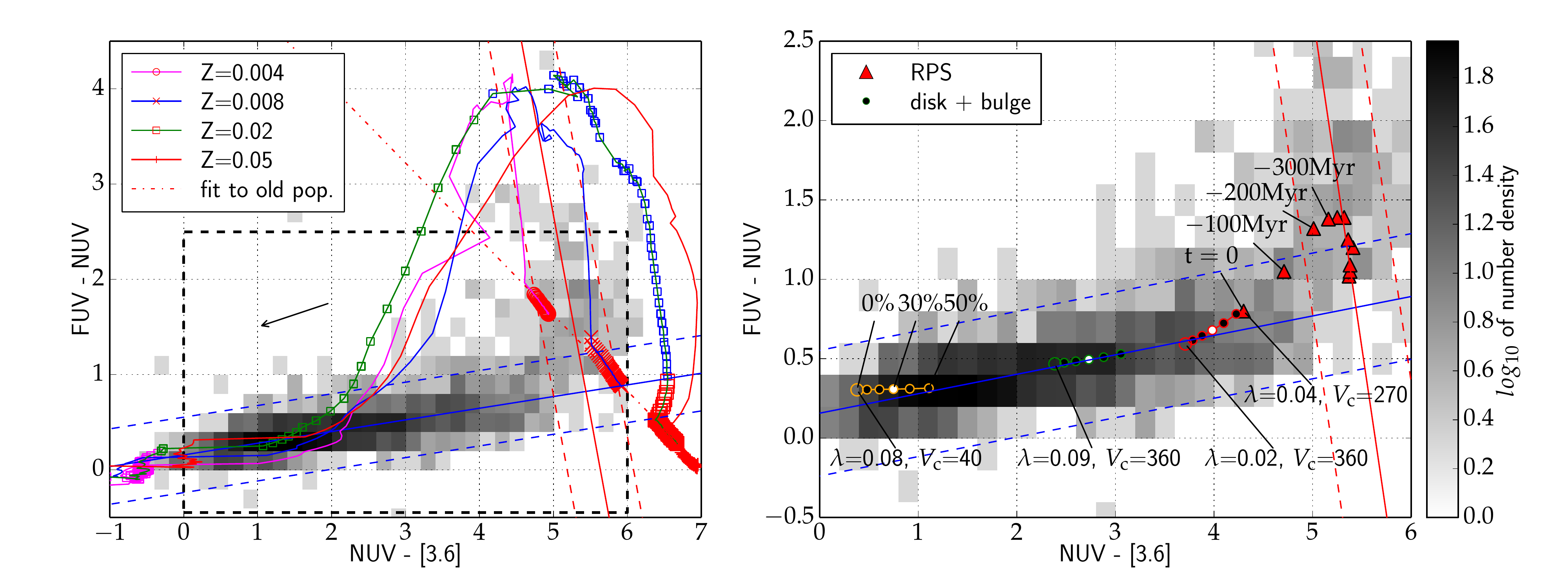}
\caption{Number density of galaxies in the ($\textit{FUV}$\,$-$\,$\textit{NUV}$) versus ($\textit{NUV}$\,$-$\,[3.6]) color-color diagram, in grayscale-coded 2D bins on a logarithmic scale (see color bar). Each bin is 0.2\,$\times$\,0.2\,mag in size. \textit{Left}\,: single stellar population (SSP) models (CB07, Chabrier IMF) for various metallicities are over-plotted on the number density distribution. The \textit{open squares} are various age time steps for the model with Solar metallicity ($\textit{Z}$=0.02), color-coded according to the following log(age) ranges: \textit{magenta} 5.0 -- 8.0, \textit{green} 8.0 -- 9.0, \textit{blue} 9.0 -- 10.0, and \textit{red} 10.0 -- 10.3. The age increases from left to right, from log(age)\,= 5.0 to 10.3, for all models. Evolved model points (log(age)\,$\geq$\, 10.0) are plotted in \textit{red} for all models. A linear fit to these points (dash-dotted line) is traced for visual comparison with the slope of the GRS. The black rectangle outlines the region of the right panel. \textit{Right}\,: The red \textit{triangles} at ($\textit{NUV}$\,$-$\,[3.6])\,$>$\,4 are model cluster galaxies under the effect of ram-pressure stripping \citep[RPS; see][]{Boselli2006} where the annotations indicate when the RPS started (with $\textit{t}$\,=\,0 being the present). The model indicate that the transition timescale of galaxies from the GBS to the GRS is very short, on the order of a few $10^{8}$\,years. We also show three examples, as \textit{orange}, \textit{green}, and \textit{red} dotted lines, of BP00\,+\,BC03 ``disk\,+\,bulge" models for various $\textit{B/T}$ ratios, where each node corresponds to a value of the $\textit{B/T}$ ratio, namely 0.0 (i.e., BP00 disk only, \textit{open circles}), 0.1, 0.2, 0.3, 0.4, and 0.5, progressively giving bluer to redder results. The position of the maximum $\textit{B/T}$\,=\,0.3 estimated by \citet{Laurikainen2007} for a sample of S0-a galaxies is shown with $\textit{white dots}$. An offset of 0.3 magnitude was added to the ($\textit{FUV}$\,$-$\,$\textit{NUV}$) colors for all model galaxies (see text). Both BP00 disk models and RPS models are based on the Kroupa 93 IMF.
\label{fig:NumDensity}}
\end{center}
\end{figure*}
In Fig.\ref{fig:Optical-IR-CMD} we show the optical CMD, commonly used to study galaxy evolution \citep[e.g.,][]{Faber2007}. We show the position of our GBS, GRS, and GGV galaxies using blue, red, and green points, respectively (Fig.\ref{fig:ColorColorDiagram}).

Despite the low number of GGV galaxies overall, they fall either on the optical red sequence (RS) or on the (ill-defined) optical green valley (GV), with only a very few being optically blue galaxies. This result indicate that GGV galaxies are evolving {\it off} the GBS much slower in ($\textit{NUV}$\,$-$\,[3.6]) than in ($\textit{FUV}$\,$-$\,$\textit{NUV}$), while the optical colors are reddened by more than [0.4\,mag] 0.2\,mag in [(U$-$B)] (B$-$V). Alternatively, the GGV galaxies are optically red systems that are growing a disk ({\it towards} the GBS), in which case the optical reddening is less than [0.3\,mag] 0.2\,mag in [(U$-$B)] (B$-$V).

It is interesting to note that the majority of the galaxies in the optical GV belong to the GBS, i.e., they are star-forming systems to be classified as such at UV wavelengths. The same can be said about the blue dots found in Fig.\ref{fig:Optical-IR-CMD} near the optical RS. This means that either they are in the tail end of star formation or they are regrowing a disk. In particular, a disk regrowth is unlikely, given that the GGV galaxies are found preferentially in dense environments where such a process may be more difficult than in the outskirts of a cluster or a lower-density environment (Section \ref{ENVIRONMENT}).

By using the classification of interacting systems within S$^{4}$G given by \citet{Knapen2014} we do not find a clear difference between the fraction of close interactions in the three regions of our color-color diagram. When splitting the sample by level of interaction and morphological type, low number statistics hamper determining, for example, whether or not E/S0 in the GGV are mergers or post-starbursts.

\section{Discussion}
\subsection{Modeling color evolution}
In Fig.\ref{fig:NumDensity} (left), we show single stellar population (SSP) models (initial mass function (IMF): Chabrier) of Charlot and Bruzual (2007) for different metallicities and ages. For all cases, the (FUV\,$-$\,NUV) color becomes redder until it reaches a peak (at about 1\,Gyr in the case of Solar metallicity), and then becomes bluer again (UV-upturn) as the models age. For these SSP models, the color transition from GBS to GRS takes less than a few 10$^{8}$\,years.

To better understand what drives the position of galaxies in the color-color diagram, we use a grid of disk galaxy models from \citet{BP00} with various circular velocities $V_{\rm{c}}$ and spin parameters $\lambda$ (Fig.\ref{fig:NumDensity}, right panel). These models span the entire GBS due to their variation in star formation timescale with $V_{\rm{c}}$ and $\lambda$. Besides, the reddening arrow points along the GBS, so internal extinction contributes little to its spread, since star formation timescale and dust attenuation are nearly degenerate in these colors (Section \ref{REDDENING}). The tightness of the GBS is also due to the fact that the effects of an episodic SFH in the FUV-NUV color (compared to the NUV-[3.6]) are only expected at very low masses (typically rather blue NUV-[3.6] colors). Nice examples in this regard are PGC065367 or ESO245-007. To produce more realistic galaxy colors, we add a bulge to the \citet{BP00} disks using SSP predictions with ages between $10^{9}$ and $10^{10.3}$\,Gyr and different formation timescale $\tau$. We adopt a bulge model with ($\textit{NUV}$\,$-$\,[3.6])\,=\,5.3\,mag, and ($\textit{FUV}$\,$-$\,$\textit{NUV}$)color\,=\,2.4\,mag.

Three of these ``disk+bulge" models are shown in Fig.\ref{fig:NumDensity} (right) for various bulge-to-total mass ratios $\textit{B/T}$. Each colored line represents the colors of the resulting galaxy model for $\textit{B/T}$ from 0.0 to 0.5 in steps of 0.1. Adding a BC03 SSP-based bulge to any of our disk models does not reproduce the reddening in (FUV\,$-$\,NUV) observed in numerous early-type spirals unless we assume a very high $\textit{B/T}$ \citep[typical observed $\textit{B/T}$ of S0 and S0-a galaxies are below 30\%;][]{Laurikainen2007}. This shows that the position of transitional early-type spirals is not only due to their (red) massive bulges but also to their disks being red compared to the disks of galaxies in the GBS. This result explains the red ($\textit{FUV}$\,$-$\,$\textit{NUV}$) colors of some (transitional) disk galaxies. Whether those disk galaxies could evolve into systems with properties similar to those of current-day ETGs cannot be determined from these data alone. %On the other hand, it is worth noting that ellipticals with a 20\% disk in stellar mass would already be outside the GRS.

To analyze the transformation timescale of these disks under a more realistic scenario, we use a model that includes ram-pressure stripping of the gas (RPS hereafter) as a potential mechanism, albeit not the only (potential) one, to redden a galaxy disk. This RPS model starts from an unperturbed disk with $V_{\rm{c}}$\,=\,270\,km\,s$^{-1}$ and $\lambda$\,=\,0.04 but with additional ram-pressure to fit the properties of the Virgo galaxy NGC 4569 \citep{Boselli2006}. The points represent the epochs when the peak of RPS occurred, with $\textit{t}$\,=\,0 corresponding to the present. Both the disk models and the peak RPS models are shifted redward by 0.3\,mag in ($\textit{FUV}$\,$-$\,$\textit{NUV}$) to match the observed color-color distribution. The need for an offset has been discussed in \citet{Munoz2011} and is believed to be due to uncertainties in the calibration of the stellar atmospheres used in the spectral synthesis. This latter RPS disk model fits the observed distribution well, especially with a moderate bulge. The transition timescale derived from the GBS to the GRS in this case would again be of the order of a few $10^{8}$\,years.

\subsection{Internal dust reddening} \label{REDDENING}
To analyze the effects of internal dust, we show the reddening in both colors in Fig.\ref{fig:ColorColorDiagram} (left), for $A_{\rm{V}}$\,=\,0.5\,mag, assuming the attenuation law of \citet{Calzetti1994}. This law reproduces well the relation between IR excess and UV color \citep{gdp2007b}. Interestingly, the reddening arrow indicates an almost complete degeneracy in these colors between reddening and a change in the star formation timescale \citep[as illustrated by the models of ][]{BP00}.
We find no change in the distribution of offsets from the GBS as a function of the galaxies' axial ratio (not shown), so the (small) spread of the GBS is unlikely to be due to inclination effects.

\subsection{Environmental effects} \label{ENVIRONMENT}
\begin{figure*}
\begin{center}
\includegraphics[width=1\textwidth]{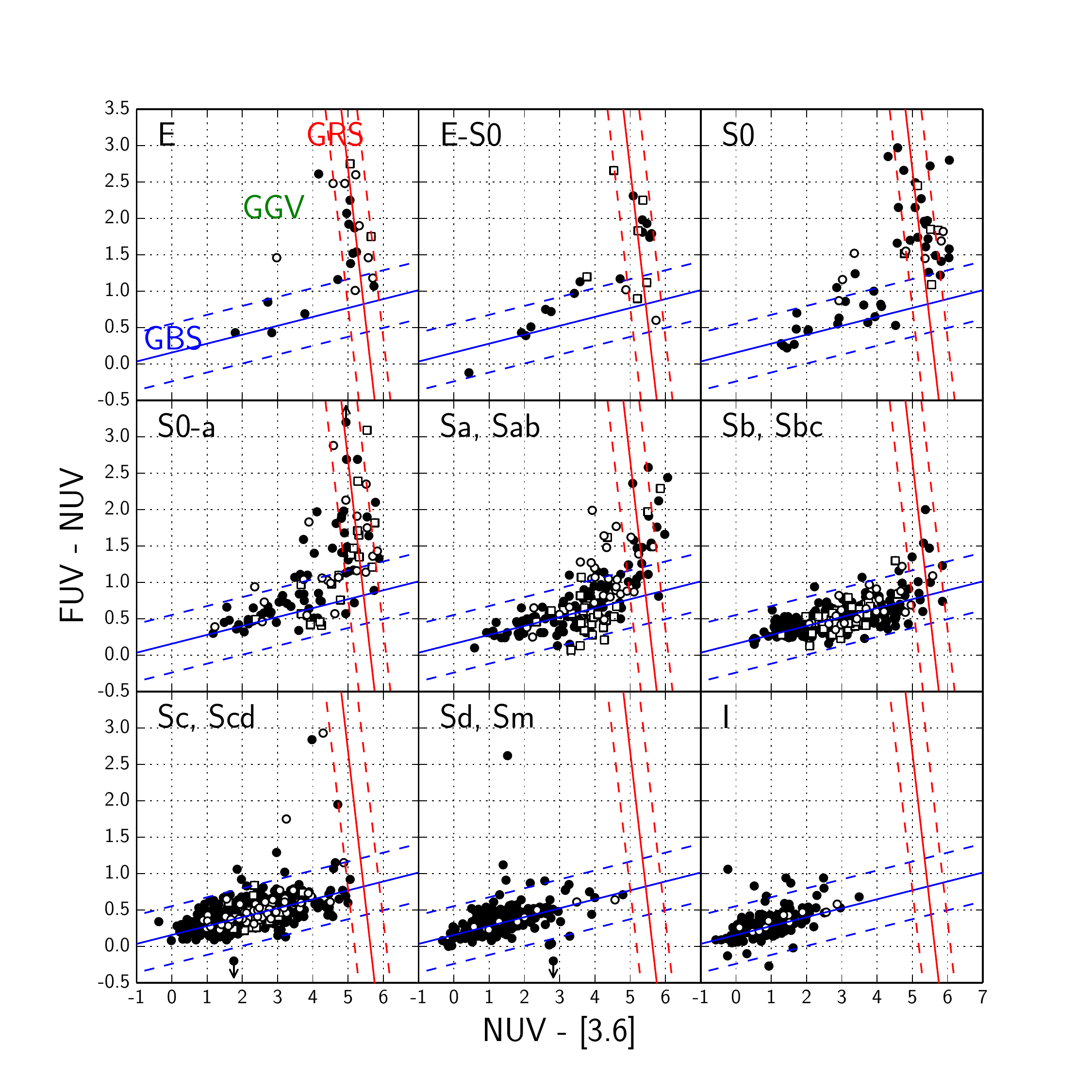}
\caption{($\textit{FUV}$\,$-$\,$\textit{NUV}$) versus ($\textit{NUV}$\,$-$\,[3.6]) separated into nine panels by morphological type. Black circles are galaxies with no environmental data associated. Open circles are galaxies in the Virgo cluster. Open squares are galaxies in high-density regions as defined by \citet{Laine2014}. The GRS, GGV, and GBS are shown in panel E.\label{fig:nine}}
\end{center}
\end{figure*}

In Fig.\ref{fig:nine} we show, superposed on the number density color-color diagram, the distribution of galaxies that are in Virgo or in high-density regions \citep{Laine2014}. The relative abundance of transitional galaxies in the Virgo cluster is higher for the majority of morphological types, suggesting that an environmental effect is present.
There are some indications that environment plays a role in determining whether galaxies lie in the GBS, GGV, or GRS. For example, 43\% (12 of 28) of Sa galaxies classified as either GRS or GGV are in the Virgo cluster, while only 12\% (16 of 137) of those classified as GBS are in Virgo. The relative preference of GGV and GRS galaxies for the Virgo environment is found for each morphological class separately (Table~\ref{table1}), although in some cases the results are not statistically significant.
%For example (see Table~\ref{table1}), 43\% of transitional Sa spirals (GRS\,+\,GGV) are in Virgo, compared to 12\% of Sa galaxies (out of 137) that are in the GBS. We note that for all types, the ratio of Virgo galaxies (162/1941\,=\,8\% of the current sample) in the GBS is 7\% (124/1759), in the GGV is 27\% (20/73), and in the GRS is 18\% (14/79), and 13\% (4/30) for the rest. Thus, the GGV has a higher fraction of Virgo galaxies than other regions. The most abundant type in the GGV is Sa (14 out of 73 GGV galaxies), where 64\% (9/14) of Sa galaxies are in Virgo. Although the presence of GRS galaxies in the optical GV supports a quenching scenario (see \citet{Boselli2008}), we cannot exclude the possibility of having the opposite, especially in the field, i.e., a transition from the GRS to the GBS.
Finally, although the presence of early-type spirals in the GGV might be related to environment, this might not be simply due to RPS, there might be other, more fundamental drivers (that also depend on environment) which determine the position of galaxies in this color-color diagram and in the optical CMDs. One possibility might be the degree of uniformity of the angular momentum of accreted material, which would segregate galaxies by morphological type, mass and color in different environments (less uniform accretion having likely occurred in denser environments).   

\section{CONCLUDING REMARKS}
We show that star-forming galaxies form a tight ``Blue Sequence" in $\textit{FUV}$-to-near-infrared color-color diagrams (the ``$\textit{GALEX}$ Blue Sequence'', or simply GBS). One could even define a star-forming galaxy (SFG) as an object residing in that sequence. This is a clear advantage over the optical or optical-IR color-magnitude and color-color diagrams where star-forming galaxies are mixed with other types of sources, such as post-starbursts or, in general, galaxies in the $\textit{GALEX}$ Green Valley (GGV). Sb, Sa, and S0 galaxies are found all across the path between the GBS and the GRS, through the GGV. Their position is not only driven by their stellar mass or their bulge mass, but also by the color evolutionary state of their disk. Both SSP and realistic disk+bulge models with RPS indicate that the timescale for the color-transition could be as fast as a few 10$^{8}$\,years. We cannot determine based solely on the integrated UV, optical and IR colors of these systems whether they are evolving off or towards the GBS.  

The fact that transitioning galaxies are more common in clusters and in high-density regions in general suggests that environment plays a role and is likely to dim their disks. Environment-related mechanisms other than RPS, such as gas starvation \citep[which can start happening in groups;][]{Kawata2008}, galaxy harassment or processes related to the degree of uniformity of accretion are all viable. There could be galaxies in the field evolving backwards, i.e., from the GRS {\it towards} the GBS, including a large number of optical GV galaxies that we find in the GBS. A future spatially-resolved analysis may provide clues to discriminate between these mechanisms, since the radial color gradients may be different from one process to another. For example, RPS should lead to strong color gradients as the low density gas in the outskirts of galaxies would suffer the RPS effects first, while star formation in the inner parts should be unaffected.  

\acknowledgments
The National Radio Astronomy Observatory is a facility of the National Science Foundation operated under cooperative agreement by Associated Universities, Inc. We acknowledge financial support to the DAGAL network from the People Programme (Marie Curie Actions) of the European Union's Seventh Framework Programme FP7/2007-2013/ under REA grant agreement number PITN-GA-2011-289313. JHK acknowledges financial support from the Spanish MINECO under grant number AYA2013-­41243-­P. AGdP and JG acknowledge financial support under grant number AYA2012-30717. 

%TABLES
%\newpage
\begin{table*}
\begin{center}
\caption{Counts and ratios of Virgo galaxies per morphological type bin per region\label{table1}}
\begin{tabular}{|c|rrr|rrr|rrr|rrr|rrr|}
\hline
%Type & S4G/GALEX & subset1 & subset2 & vall & vsub1 & vsub2 &
%\multicolumn{1}{c}{$ratio all$\tablenotemark{a}} & ratio sub1 & ratio sub2\\
&\multicolumn{3}{ |c| }{GBS}&\multicolumn{3}{ |c| }{GGV}&\multicolumn{3}{ |c| }{GRS}&\multicolumn{3}{ |c| }{Others\footnotemark[1]}&\multicolumn{3}{ |c| }{ALL}\\
\hline
Type\footnotemark[2]&Virgo\footnotemark[3]&all\footnotemark[4]&ratio\footnotemark[5]&Virgo&all&ratio&Virgo&all&ratio&Virgo&all&ratio&Virgo&all&ratio\\
\hline
E&2&7&0.29&2&4&0.50&4&13&0.31&---&---&---&8&24&0.33\\
E-S0&2&11&0.18&---&4&---&---&8&---&---&---&---&2&23&0.09\\
S0&1&20&0.05&3&9&0.33&1&15&0.07&2&7&0.33&7&51&0.14\\
S0-a&11&61&0.18&3&12&0.25&6&24&0.25&2&6&0.22&22&103&0.21\\
Sa&16&137&0.12&9&14&\textbf{0.64}&3&14&0.21&---&10&---&28&175&0.16\\
Sb&21&331&0.06&1&5&0.20&---&4&---&---&---&---&22&340&0.06\\
Sc&51&656&0.08&2&11&0.18&---&1&---&---&1&---&53&669&0.08\\
Sd&4&162&0.02&---&3&---&---&---&---&---&3&---&4&168&0.02\\
Sm&5&189&0.03&---&2&---&---&---&---&---&1&---&5&192&0.03\\
Irr&11&179&0.06&---&6&---&---&---&---&---&1&---&11&186&0.06\\
\hline
ALL types&124&1753&0.07&20&70&0.29&14&79&0.18&4&29&0.14&162&1931&0.08\\
\hline
\end{tabular}
\end{center}
%% Any table notes must follow the \end{tabular} command.
\footnotemark[1]{Galaxies where ($\textit{FUV}$\,$-$\,$\textit{NUV}$)$<$\,2\,$\sigma$ of the GBS fit and ($\textit{NUV}$\,$-$\,[3.6])\,$>$\,1$\sigma$ of the GRS fit.}\\
\footnotemark[2]{Based on Hubble t-type from RC2}\\
\footnotemark[3]{From the GOLDMine database \citep{Gavazzi2003}}\\
\footnotemark[4]{including Virgo galaxies}\\
\footnotemark[5]{ratio=Virgo/all}
%\tablecomments{We can also attach a long-ish paragraph of explanatory
%material to a table.}
\end{table*}

\bibliographystyle{apj}	
%\bibliography{dagalbib.bib}		% expects file "myrefs.bib"

\end{document}